\documentstyle[11pt]{article}
\def\thefootnote{\fnsymbol{footnote}}
\def\bea {\begin{eqnarray}}
\def\eea {\end{eqnarray}}
\def\be {\begin{equation}}
\def\ee {\end{equation}}
\def\ben{\begin{enumerate}}
\def\een{\end{enumerate}}
\def\bi{\begin{itemize}}
\def\ei{\end{itemize}}
\def\ie{{\it i.e.}}
\def\viz{{\it viz}}

\def\etal{{\it et al.}}
\def\rhat{{\hat r}}
\def\prl {Phys. Rev. Lett.}
\def\pl {Phys. Lett.}
\def\pr {Phys. Rev.}
\def\np {Nucl. Phys.}
\def\F{f_\pi}
\def\gA{g_{\! A}}
\def\mN{m_{\mbox{\tiny N}}}
\def\kF{k_{\mbox{\tiny F}}}
\def\M{{\hat {\cal M}}}
\def\T{{\hat {\cal T}}}
\def\hyphen{{\mbox{-}}}
\def\smallminus{\!-\!}
\newcommand{\ff}[3]{{#1}\mbox{#2}_{{#3}/2}}
\newcommand{\Aeq}[1]{{\mbox{$A$={#1}}}}

\newcommand{\rmbody}[1]{{{#1}\hyphen\rm{body}}}
\def\Ox{{}^{16}\mbox{O}}
\def\Sn{{}^{132}\mbox{Sn}}
\def\Pb{{}^{208}\mbox{Pb}}

\begin{document}
\begin{titlepage}
\begin{center}
\begin{flushright}
{SNUTP 94-29, USC(NT)-94-2\\ nucl-th/9404015}
\end{flushright}
\vskip 1.0cm
{\large\bf Nuclear Matrix Elements of Axial-Charge Exchange
Currents}
\vskip 0.2cm
{\large\bf  Derived in Heavy-Fermion Chiral Perturbation Theory}
\vskip 2cm
{\large Tae-Sun Park}$^{\S \, \ddag}$\\
{\it Department of Physics and Center for Theoretical Physics}\\
{\it Seoul National University}\\
{\it Seoul 151-742, Korea}
\vskip 0.4cm
{\large Ian S. Towner}\\
{\it AECL Research, Chalk River Laboratories, Chalk River}\\
{\it Ontario K0J 1J0, Canada}
\vskip 0.4cm
{\large Kuniharu Kubodera}$^{\ddag}$\\
{\it Department of Physics and Astronomy}\\
{\it University of South Carolina}\\
{\it Columbia, SC 29208, U.S.A.}
\vskip 2cm

{\bf Abstract}
\end{center}

\begin{quotation}
We calculate shell-model matrix elements
of the axial-charge exchange current operators
that have been obtained up to the next-to-leading order
from heavy-fermion chiral perturbation theory.
It is found that loop corrections
to the soft one-pion-exchange contribution
are small (around 10 \%)
and have no significant dependence
on the nuclear mass number
or on the valence-nucleon orbits.
These results render further support
to the chiral-filtering conjecture.

\end{quotation}

\vfill
\noindent
$^{\S}$Supported in part by the Korea Science
and Engineering Foundation
through the Center for Theoretical Physics,
Seoul National University.

\noindent
$^{\ddag}$Supported in part by the National Science Foundation
(USA), Grant No. PHYS-9310124.

\end{titlepage}
\renewcommand{\thefootnote}{\#\arabic{footnote}}
\setcounter{footnote}{0}
\section{Introduction}
\indent \indent

The nuclear axial-charge operator,
which carries valuable information about
the role of the chiral symmetry in nuclear medium,
has been the subject of
a great deal of experimental and theoretical work
(for reviews, see \cite{tow86,kub93}).
The crucial point is that $A^0$,
the time component of the axial current,
is predicted to have a large exchange current due to
a one-pion exchange diagram,
the structure and strength of which are dictated by
the soft-pion theorem \cite{kdr78}.
According to a general argument
by Kubodera, Delorme and Rho \cite{kdr78},
the impulse-approximation (IA)
one-body current $A^0(1\hyphen{\rm body})$
should receive a large meson-exchange-current correction
$A^0({\rm mec})$
which is mainly due to the soft-pion-exchange
two-body current $A^0({\rm soft}\hyphen\pi)$:
\be
A^0 = A^0(1\hyphen{\rm body}) + A^0({\rm mec})
\ee
where
\bea
 A^0(1\hyphen{\rm body}) = \frac{\gA}{2\mN} \,
 \vec{\sigma} \cdot \vec{p} \,\,
 \tau^{\pm},
\label{a1}
\eea
\bea
 A^0({\rm{mec}}) \simeq
 A^0({\rm soft}\hyphen\pi) =
 \frac{\gA}{4{\F}^2}
 \frac{m_\pi}{4\pi r}
   \left(1 + \frac{1}{m_\pi r}\right)
   {\mbox{e}}^{- m_\pi r}\,\,
    (\vec{\sigma}_1 +\vec{\sigma}_2)\cdot\hat{r}\,
 (\vec{\tau}_1 \times \vec{\tau}_2)^{\pm},
 \label{a2}
\eea
with $\vec{r}=\vec{r}_1-\vec{r}_2$.
Here,
${\vec p}=\frac12 ({\vec p}_i+{\vec p}_f)$ with ${\vec p}_i$
and ${\vec p}_f$ the incoming and outgoing nucleon momentum;
$\F$ is the pion-decay constant, $\gA$
the axial-vector coupling constant, $\mN$ the nucleon mass and
$m_\pi$ the pion mass.
A useful measure of
the typical effects of $A^0({\rm soft}\hyphen\pi)$
was given by Delorme \cite{del82}.
For a system of one valence nucleon added to the Fermi gas,
Delorme derived the Hartree-Fock-type
effective single-particle operator
arising from $A^0({\rm soft}\hyphen\pi)$,
and evaluated its strength
relative to the single-particle operator, eq.(\ref{a1}).
This Fermi gas model estimate indicates that
the original single-particle operator should
be enhanced by $\sim 54 \%$ ($\sim 39 \%$)
at nuclear matter density (half nuclear density).
Thus, indeed, the two-body effect is very large.
The replacement of
the uncorrelated two-nucleon relative-motion
wave functions used in the Fermi gas model
with typical short-range-correlated wave functions
would reduce the enhancement
to some extent but,
even with this reduction,
the effect of $A^0({\rm soft}\hyphen\pi)$
is still expected to be of a substantial magnitude.

All the existing experimental information
supports the significant enhancement of
$ A^0(1\hyphen{\rm body})$
due to the soft-pion exchange current
(see the extensive bibliographies
in \cite{tow86,kub93}).
The most convincing evidence
comes from Warburton \etal's systematic analyses
\cite{War,WIAN}
of the first-forbidden transitions
over a wide range of the periodic table.
The latest results of these analyses \cite{WIAN}
can be summarized in terms of $\delta_{\rm{mec}}$,
the ratio of the meson-exchange-current contribution
to the IA one-body contribution:
$$\delta_{\rm{mec}} \, = \,
\frac{\langle A^0({\rm mec})\rangle}{
\langle\!A^0(1\hyphen{\rm body})\rangle } .$$
The best fit to data gives:
\bea
\delta_{\rm{mec}}(\Ox)  &=& 0.61(3),\nonumber \\
\delta_{\rm{mec}}(\Sn)  &=& 0.82(7),\nonumber \\
\delta_{\rm{mec}}(\Pb)  &=& 0.79(4).
\eea
These results on $\delta_{\rm{mec}}$
are certainly compatible with
the sizeable contribution of $A^0({\rm soft}\hyphen\pi)$;
in fact, the empirically determined enhancements
even seem somewhat stronger than anticipated from
the $A^0({\rm soft}\hyphen\pi)$ contribution.

Recent theoretical developments
are concerned with:
(i) the evaluation of heavy-meson exchange diagrams
\cite{krt92,A542};
(ii) the estimation of higher order terms
in chiral perturbation \cite{r91,PMR};
(iii) possible consequences of the in-medium scaling
of particle masses \cite{kr}.

All these developments are related
to the chiral-filtering conjecture \cite{kdr78,rb81},
which states that,
whenever there is no kinematical suppression,
one-soft-pion exchange currents should dominate,
with other shorter-ranged currents ``filtered" off
by the nuclear medium.
Although a reasonably reliable argument
for the dominance of $ A^0({\rm soft}\hyphen\pi)$
in $A^0(2\hyphen{\rm body})$ was presented in \cite{kdr78},
the explicit demonstration of
the suppression of other short-ranged contributions
was left for future studies.
The subsequent accumulation of
experimental evidence in support of
the chiral-filtering mechanism
has further increased the importance
of this explicit demonstration.

Several authors addressed this and related issues
by evaluating the Feynman diagrams involving
heavy-meson exchange \cite{krt92,A542}.
The upshot of this approach is that
some of the short-range exchange currents can be
rather large but that they tend to cancel among themselves.
The net contribution of the heavy-meson exchange diagrams
is to enhance $A^0({\rm soft}\hyphen\pi)$
by about the right amount to reproduce $\delta_{\rm{mec}}$.
Although the idea of heavy-meson-exchange currents
emerges naturally from
the picture of the boson-exchange nuclear interactions
and from the Dirac phenomenology \cite{wal74,DT87}
and although its numerical results are generally encouraging,
the convergence property of this approach
is a non-trivial question.
At present, there seems to be no clue to
a relevant expansion parameter
that controls a multitude of complicated diagrams
describing short-range exchange currents.
Also, the drastic cancellation
between the individually large terms
tends to make their net contribution
quite model-dependent,
a problem which awaits further investigations.

A systematic study of $A^0(2\hyphen{\rm body})$
based on chiral perturbation theory (ChPT)
was taken up by Rho \cite{r91}
with the use of the heavy-fermion formalism (HFF)
\cite{weinberg,jm91,georgi}.
In ChPT,
the chiral symmetry and well-defined counting rules
allow a systematic construction of
an effective Lagrangian.
Furthermore, the use of HFF enables us to
carry out a consistent perturbation expansion
even for systems that involve the nucleons\footnote{
{}From a formal point of view
the general application of ChPT to a many-body system
involves a subtlety,
since the system has an extra scale
$k_{\rm{F}}$ (Fermi momentum)
in addition to the usual chiral-symmetry breaking scale
$\Lambda \sim $1 GeV.
However, it seems justifiable to use ChPT
in the following limited context.
We start with the generally accepted paradigm
that the diluteness of nuclear matter
allows us to concentrate on 1-body and 2-body responses
to external probes, ignoring $n$-body ($n \geq 3$) contributions,
and that the effective operators
describing these responses can be obtained
by considering Feynman diagrams involving
one nucleon (two nucleons) for the 1-body (2-body) operators.
Once a one-nucleon or two-nucleon subsystem
is isolated from the $A$-body system,
we use ChPT as a method
to calculate the Feynman diagrams pertaining to these subsystems
systematically and
consistently with the basic chiral symmetry of QCD.
In the present paper, the word ``ChPT" should be
understood in this context.}.
A proof of the soft-pion-exchange dominance
to leading order of chiral expansion was given
by Rho \cite{r91}.
The extension of the ChPT calculation
to the next order (one-loop order)
was carried out by Park \etal\ \cite{PMR}.
Park \etal's calculation
indicates that the loop correction
to the soft-pion-exchange contribution is very small,
supporting the robustness of the chiral filter mechanism.
In HFF, baryon momentum-dependent terms appear
as higher order interaction terms in the chiral counting,
and this feature leads to
a drastic reduction of the contributions
of the heavy-meson exchange diagrams.
The relation between the heavy-meson-exchange approach
and ChPT is an interesting problem that warrants
further careful studies.

In \cite{PMR},
the axial-charge transition matrix elements
were calculated with the simple-minded Fermi-gas model.
To obtain more reliable estimates of
the contributions of higher-order terms in the
chiral expansion,
we need to use more realistic nuclear wave functions
than those of the Fermi gas.
Also, according to Towner \cite{A542},
the matrix elements of the heavy-meson exchange currents
exhibit significant dependence on the valence orbits.
It is important to examine
whether or not the loop corrections in ChPT
shows a similar shell dependence.
In this note, we calculate the shell-model
matrix elements of
the axial-charge exchange current operators
obtained in \cite{PMR}
up to next-to-the-leading order in ChPT.
Our results indicate that
the loop corrections calculated in the shell model
are as small as indicated by the Fermi gas model estimation.
Furthermore, the mass- and state-dependence
of the loop correction is negligible.
These results give further support to
the chiral-filtering conjecture.
In the light of our new results,
we shall also discuss the problem of
the ``extra" enhancement in the empirical
$\delta_{\rm{mec}}$.
In particular, we comment
on the interrelation between
the heavy-meson-exchange-current method
\cite{krt92,A542} and
the ChPT with in-medium mass scaling \cite{kr}.

\section{Calculation}
\indent \indent
We begin by specifying our conventions and definitions.
We introduce, as we did in \cite{PMR}, the quantity $\M$,
which is different from the axial-charge operator $A^0$
by just an overall factor,
\be
\M = \frac{4\F^2}{\gA}\, A^0 = \M_\rmbody{1} +
\M_\rmbody{2} + \cdots . \label{eq:a}
\ee
Using $\M$ is slightly more convenient
than using $A^0$ itself.
In eq.(\ref{eq:a}), $\M_\rmbody{n}$ represents
an $n$-body operator, and
the ellipsis denotes $\M_\rmbody{n}$ ($n \geq 3$),
which are ignored.
$\M_\rmbody{1}$ is the impulse approximation term,
\be
\M_\rmbody{1} = \frac{2 \F^2}{\mN}\,
{\vec \sigma}\cdot
{\vec p}\, { \tau}^{\pm}.
\ee
The two-body exchange current,
$\M_\rmbody{2}$, is decomposed into a tree part
and a loop correction.
The tree part is the seagull term considered in
the original paper \cite{kdr78}.
Meanwhile according to \cite{PMR},
the loop correction can be decomposed
into a one-pion exchange part
(a loop correction to the tree part through vertex
renormalization)
and a two-pion-exchange part :
\bea
\M_\rmbody{2} &=& \M_{\rm{tree}} + \M_{\rm{loop}},
\nonumber \\
\M_{\rm{loop}} &=& \M_{1\pi} + \M_{2\pi},
\eea
with
\bea
\M_{\rm{tree}} &=& \T^{(1)}\, f_{\rm{tree}}(r)
\nonumber\\
\M_{1\pi} &=& \T^{(1)}\,f_{1\pi}(r),
\nonumber \\
\M_{2\pi} &=& \T^{(1)}\, f_{2\pi(1)}(r) + \T^{(2)}\, f_{2\pi(2)}(r).
\label{twobd}
\eea
The spin-isospin operators $\T^{(1)}$ and $\T^{(2)}$ are defined as
\bea
\T^{(1)}= ({\vec \tau}_1\times {\vec \tau}_2)^{\pm} \,\,
 \rhat\cdot({\vec \sigma}_1 + {\vec \sigma}_2),
\nonumber \\
\T^{(2)}= ({\vec \tau}_1 + {\vec \tau}_2)^{\pm} \,\,
 \rhat\cdot({\vec \sigma}_1 \times {\vec \sigma}_2),
\eea
where $\rhat$ is the unit vector of ${\vec r}= {\vec r}_1 - {\vec
r}_2$
and $r$ is the norm of ${\vec r}$.  Functions $f_{\rm{tree}}(r)$,
$f_{1\pi}(r)$ and $f_{2\pi(1,2)}(r)$ are given in \cite{PMR},
\bea
f_{\rm{tree}}(r) &=& \frac{d}{dr}\left(
- \frac{1}{4 \pi r} {\mbox{e}}^{- m_\pi r}\right),
\label{msoft} \\
f_{1\pi}(r) &=& c_3^R \frac{m_\pi^2}{\F^2}\,f_{\rm{tree}}(r)
\nonumber \\\
&+&\frac{1}{16\pi^2 \F^2}
\frac{d}{dr} \left(
 - \frac{1+3 \gA^2}{2}\left[K_0(r)- {\tilde K_0}(r)\right]
 + (2+4\gA^2) \left[K_2(r) - {\tilde K_2}(r)
\right]\right),\nonumber\\
\label{onepiloop}\\
f_{2\pi(1)}(r) &=& \frac{1}{16\pi^2 \F^2} \frac{d}{dr}\left(
-\frac{3\gA^2-2}{4} K_0(r) -\frac12 \gA^2 K_1(r) \right),
\nonumber \\
f_{2\pi(2)}(r) &=& \frac{1}{16\pi^2 \F^2} \frac{d}{dr}\left(
2\gA^2 K_0(r)\right),
\label{twopiloop}
\eea
The explicit forms of the functions $K_i(r)$ and ${\tilde K}_i(r)$
are given in Appendix B of ref.\cite{PMR}.

As noted in \cite{PMR},
the constant $c_3^R$ can be extracted from the isovector
Dirac form factor of the nucleon, \ie,
\be
c_3^R \frac{m_\pi^2}{\F^2} = \frac{m_\pi^2}{6} \langle r^2\rangle_1^V
\simeq 0.04784.\label{c3r}
\ee

In shell-model calculations, it is convenient to write the above
expressions in terms of spherical tensors \cite{A540},
\bea
i \M_\rmbody{1} &=& -\sqrt{3}\,\,\frac{2 \F^2}{\mN}\,{
\tau}^{\pm}
\, \left[{\vec \sigma},\,\, i {\vec p}\,\right]^{(0)},
\nonumber \\
i \M_i &=& -\sqrt{6}\,\, \Sigma^{(1)}_m(\tau)\,\left[\rhat,\,
\Sigma^{(1)}_n(\sigma) \right]^{(0)}\, f_i(r)
\eea
where $(m,n)= (2,4)$ for $i= \mbox{`tree'},
\ 1\pi,\ 2\pi(1)$, which
are proportional to $\T^{(1)}$, and $(m,n)=(4,2)$ for $i=2\pi(2)$,
which is proportional to $\T^{(2)}$.  Here, $\Sigma^{(1)}_m(\tau)$
and
$\Sigma^{(1)}_n(\sigma)$ are defined in \cite{A540},
\bea
\Sigma^{(1)}_2(\tau) &=&\frac{i}{\sqrt{2}}\,({\vec \tau}_1 \times
{\vec \tau}_2)^{\pm}, \nonumber \\
\Sigma^{(1)}_4(\tau) &=& ({\vec \tau}_1 + {\vec \tau}_2)^{\pm}
\eea
and similarly for $\Sigma^{(1)}_n(\sigma)$, $n=2,4$. The square
bracket
with superscript $(0)$ in the above equations represents the vector
coupling of two spherical tensors to a resultant\footnote{In general,
$$\left[Y^{(k_1)}(1),\ Y^{(k_2)}(2)\right]^{(K)}_M \equiv
\sum_{m_1, m_2} \langle k_1 m_1, k_2 m_2 | K  M\rangle\,
Y^{(k_1)}_{m_1}(1)\, Y^{(k_2)}_{m_2}(2)$$
for any tensors $Y^{(k_1)}$ and $Y^{(k_2)}$.}; in this case two vectors
are coupled to form a tensor of rank zero :
\be
\left[{\vec a}, \, {\vec b}\right]^{(0)} = -\frac{1}{\sqrt{3}} \,
{\vec a}\cdot {\vec b}.
\ee

In the simplest version of the shell model,
the ground state (Fock state) of a nucleus is taken as a Slater
determinant of single-particle states,
which here will be chosen as
eigenstates of a spherical simple harmonic oscillator.
In an oscillator Hamiltonian,
there are two parameters, one is the mass,
which will be identified as the nucleon mass,
and the other is the oscillator frequency, $\omega$,
which is determined by
\be
\hbar \omega = \left(45 \, A^{-1/3} - 25 \, A^{-2/3} \right)\,
\left[\mbox{MeV}\right]
\ee
where $A$ is the mass number of the nucleus.
For an odd-mass nucleus, the wavefunction is written as a simple
product of a Fock state for the neighbouring even-mass nucleus and
a single-particle state, $\psi_a$, \viz.
$|\psi_a ; F\rangle$.
Here the $\psi_a$ denotes
all the quantum numbers of the single-particle states
indexed by the subscript $a$,
$\psi_a= (n_a,\ l_a,\ s_a= \frac12,\ j_a,\
j_{az},\ t_a=\frac12,\ t_{az})$,
where $n_a (= 0,1,2,\cdots)$ is the principal quantum number,
$l_a$ the orbital angular momentum, $s_a$ the spin,
$j_a$ the total spin, and the subscript $z$ identifies the
third component.  The use of such a shell-model calculation is
given in detail in \cite{A540,A542}; here we just explain the
definition of reduced matrix elements and some features of our
calculation.

Consider an operator with definite multipolarity $J$ and isospin
$T$ with third component $J_z$ and $T_z$,
$X^{(J,T)}_{J_z, T_z}$; our axial-charge operator can be regarded as
$X^{(0,1)}_{0,T_z}$. Our goal is to calculate the matrix element
$\langle \psi_a; F | X^{(J,T)}_{J_z, T_z}| \psi_b; F\rangle$.
For the case when $X$ is a one-body operator,
the result trivially reduces to a single-particle matrix element
\be
\langle \psi_a | X^{(J,T)}_{J_z, T_z} |\psi_b\rangle
\equiv \langle j_b j_{bz}, J J_z | j_a j_{az} \rangle
\langle t_b t_{bz}, T T_z | t_a t_{az} \rangle
\langle a || X^{(J,T)} || b \rangle
\ee
where $a$ denotes quantum numbers of states $\psi_a$ without
third components,
$a=(n_a, l_a, j_a, s_a, t_a)$.
This expression serves to define our conventions for
reduced matrix elements.
For a 2-body operator, $X$, the reduced matrix element,
$\langle a;F || X^{(J,T)} || b;F \rangle$, is
\bea
\langle \psi_a; F | X^{(J,T)}_{J_z, T_z}| \psi_b; F\rangle
&=& \sum_{\psi_h\in F} \langle \psi_a, \psi_h| X^{(J,T)}_{J_z, T_z} |
\psi_b,\psi_h\rangle_{\mbox{\tiny AS}}
\nonumber \\
&\equiv& \langle j_b j_{bz}, J J_z | j_a j_{az} \rangle
\langle t_b t_{bz}, T T_z | t_a t_{az} \rangle
\langle a;F || X^{(J,T)} || b;F \rangle
\nonumber\eea
with
\bea
\langle a;F || X^{(J,T)} || b;F \rangle &=&
\sum_{h\in F} \sum_{T_1 T_2} \sum_{J_1 J_2} D(j_a j_b j_h,J_1 J_2,J;
F)
D(t_a t_b t_h, T_1 T_2,T;F)
\nonumber \\
&\times& \langle (ah) J_1 T_1 || X^{(J, T)} || (bh) J_2 T_2
\rangle_{\mbox{\tiny AS}}\,,
\label{reduce}\eea

\be
\langle (ah)  J_1 T_1 || X^{(J, T)} || (bh) J_2 T_2
\rangle_{\mbox{\tiny AS}} \equiv \frac{
\langle a,h ; J_1 J_{1z}, T_1 T_{1z}| X^{(J, T)}_{J_z,T_z} |
b,h; J_2 J_{2z}, T_2 T_{2z}\rangle_{\mbox{\tiny AS}} }{
\langle J_2 J_{2z}, J J_z | J_1 J_{1z}\rangle
\langle T_2 T_{2z}, T T_z | T_1 T_{1z}\rangle}
\ee
and
\be
D(j_a j_b j_h, J_1 J_2,J; F) = \sum_{j_{hz}\in F} \sum_{J_{1z} J_{2z}}
\!
\frac{\langle j_a j_{az}, j_h j_{hz}| J_1 J_{1z}\rangle
\langle j_b j_{bz}, j_h j_{hz} |J_2 J_{2z}\rangle
\langle J_2 J_{2z}, J J_z | J_1 J_{1z}\rangle}{
\langle j_b j_{bz}, J J_z | j_a j_{az}\rangle}.
\nonumber \\
\ee
A similar definition follows
for $D(t_a t_b t_h, T_1 T_2,T; F)$. The subscript `AS'
implies the matrix element is antisymmetric under the
permutation of the two-particle states.
The difference between summation over $\psi_h \in F$ and that over
$h\in F$ should be understood,
$$ \sum_{\psi_h \in F} = \sum_{h\in F}\,\, \sum_{j_{hz}\in F}\,
\sum_{t_{hz}\in F}.  $$

When the neighbouring even-mass nucleus forms a
closed shell, $D$ becomes simpler :
\be
D(j_a j_b j_h, J_1 J_2, J;F)_{\mbox{closed}}
= \frac{\sqrt{2 J_1 +1}\, \sqrt{2 J_2 +1}}{\sqrt{2 j_a + 1}\,
\sqrt{2 j_b + 1}} \, U(j_a j_h \, J J_2; J_1 j_b)
\ee
where the subscript `closed' means that $j_{hz}$ runs from $-j_h$ to
$j_h$, and $U$ is a recoupling coefficient defined in \cite{A540} and
is
related to the 6-$j$ symbol.  Now in our case,
$J=0,\ T=1,\ t_a=t_b=t_h= \frac12,\ s_a=s_b=s_h=\frac12$, and
we consider only closed shells in $j$-space\footnote{
This is not the
case for $\Aeq{96}$, where there is one unclosed shell.
Here we simply
neglect contribution coming from this unclosed shell.},
in which case we have a much simpler expression,
\be
D(j_a j_b j_h, J_1 J_2, J=0;F)_{\mbox{closed}}
= \delta_{J_1, J_2}\, \delta_{j_a, j_b}\, \frac{2 J_1 +1}{2 j_a + 1}.
\label{DJ}\ee
But in isospin space, we consider not only
charge-symmetric $N=Z$ cases but also
neutron-rich cases, and therefore we should retain the original
definition,
\bea
D_T(T_1,T_2; F) &\equiv& D(
\mbox{\small{$\frac12$}}
\mbox{\small{$\frac12$}}
\mbox{\small{$\frac12$}},
 \,T_1 T_2, T=1;F)
\nonumber \\
&=& \sum_{t_{hz}\in F} \sum_{T_{1z} T_{2z}}
\frac{\langle \frac12 t_{az}, \frac12 t_{hz}| T_1 T_{1z}\rangle
\langle \frac12 t_{bz}, \frac12 t_{hz} |J_2 J_{2z}\rangle
\langle T_2 T_{2z}, 1 T_z | T_1 T_{1z}\rangle }{
\langle \frac12 t_{bz}, 1 T_z | \frac12 t_{az}\rangle }.
\label{DT}\eea

With eqs.(\ref{reduce}, \ref{DJ}, \ref{DT}) here and eqs.(53-56) of
ref.\cite{A540}, it is straightforward to calculate the reduced matrix
elements of the axial-charge operator.

\section{Numerical Results}
\indent \indent
It is convenient to represent the results in terms of a ratio
\be
\delta_x \equiv \frac{\langle \psi_a;F| \M_x | \psi_b;
F\rangle}{\langle \psi_a; F|\M_\rmbody{1}
|\psi_b;F\rangle} =
\frac{\langle a;F|| \M_x || b; F\rangle}
{\langle a||\M_\rmbody{1}
||b\rangle} , \label{eq:deltax}
\ee
where $x=$ `tree', $1\pi,\ 2\pi$, `loop' or `2-body'.
Thus $\delta_x$ is the contribution of the $x$ part
relative to the 1-body contribution (impulse approximation):
$\delta_{\rm{tree}}$ is the ratio of
the one-pion-exchange contribution evaluated in the
soft-pion limit with no loop correction,
$\delta_{\rm{1\pi}}$ shows
the form-factor effect of one-pion-exchange
coming from the loop correction to the vertices,
and $\delta_{\rm{2\pi}}$ gives
the two-pion-exchange contribution.
The $\delta_{\rm{loop}}$
stands for the total loop contribution
compared to the impulse approximation:
\be
\delta_{\rm{loop}}= \delta_{\rm{1\pi}}
+ \delta_{\rm{2\pi}},
\ee
while $\delta_\rmbody{2}$,
which is also denoted by $\delta_{\rm{mec}}$,
represents the total 2-body contribution:
\be
\delta_{\rm{mec}}\equiv \delta_\rmbody{2}=
\delta_{\rm{tree}}+\delta_{\rm{loop}} .\label{eq:deltamec}
\ee
Furthermore, we introduce $\eta_{\rm{loop}}$
as the ratio of the total loop contribution
to the {\em tree} contribution:
\be
\eta_{\rm{loop}} \equiv
\delta_{\rm{loop}} /\delta_{\rm{tree}}
\ee

\begin{table}[t]
\begin{center}
\caption{Parameters used in the present work
\label{para}}
\vskip 1mm
\begin{tabular}{|c|c|c|}\hline
& PMR & OBEPR \\ \hline\hline
$f_\pi$ & $93.0000$ MeV & $86.3982$ MeV
\\ \hline
$m_\pi$ & $139.5673$ MeV & $138.0300$ MeV
\\ \hline
$\mN$ & $938.2796$ MeV & $938.2796$ MeV
\\ \hline
$\gA$ & $1.26$ & $1.26$\\ \hline
$d_1$ & 0.70693 fm & 0.71480 fm \\
$d_2$ & 0.5 fm  & 0.5 fm\\
$q_c$ & 3.93 fm${}^{-1}$ & 3.93 fm${}^{-1}$\\ \hline
\end{tabular}
\end{center}
\end{table}

In Table \ref{para},
we give two sets of parameters used in this
calculation; ``PMR" denotes parameters adopted in \cite{PMR},
and ``OBEPR" denotes those obtained from the
Bonn OBEPR potential \cite{OBEPR}.
There is a small difference
of $1.1$ \% in the pion mass between the two.
There is also a larger difference in the pion-decay
constant $f_\pi$: In PMR, an experimental value is adopted, while in
using OBEPR, $f_\pi$ is taken from the Goldberger-Treiman relation
and
the fitted value of the $\pi N\!N$ coupling constant of
$g_{\pi N\!N}= 13.68353$.
We observe this difference in $f_\pi$ has a significant effect on
$\delta_{\rm{mec}}$
but a smaller effect on $\eta_{\rm{loop}}$

For each parameter set, we have used three choices of
short-range correlation functions;
the first two are a simple step function with different cut-offs,
${\hat g}_i(r) = \theta(r \smallminus d_i)$,
$d_1= \frac{1}{2}  m_\pi^{-1} \simeq 0.71$ fm, $d_2= 0.5$ fm.
The third correlation function is
${\hat g}_3(r) = 1 \smallminus j_0(q_c\, r)$,
where $j_0(x)= \sin(x)/x$ and $q_c$
is fixed as $q_c= 3.93$ fm${}^{-1}$ \cite{NPA286}.

\begin{table}[t]
\begin{center}
\caption{Contributions to $\delta_{\rm{mec}}$
for various choices of parameters
for the $\ff{1}{s}{1} \rightarrow \ff{0}{p}{1}$ transition
in $\Aeq{16}$ :
${\hat g}_1(r)= \theta(r\smallminus d_1)$
with $d_1= 0.5\ m_\pi^{-1}\,\simeq\,0.71$ fm,
${\hat g}_2(r)= \theta(r\smallminus d_2)$ with $d_2= 0.5$ fm
and ${\hat g}_3(r)= 1\smallminus j_0(q_c r)$ with
$q_c= 3.93$ fm${}^{-1}$
\label{contr}}
\vskip 1mm
\begin{tabular}{|c||r|r|r||r|r|r|}\hline
& \multicolumn{3}{c|}{PMR} &
\multicolumn{3}{c|}{OBEPR}
\\ \hline\hline
${\hat g}(r)$
&${\hat g}_1(r)$ &${\hat g}_2(r)$ &${\hat g}_3(r)$
&${\hat g}_1(r)$ &${\hat g}_2(r)$ &${\hat g}_3(r)$ \\
\hline\hline
$\delta_{\rm{tree}}$  & $0.407$  & $0.459$  & $0.480$
          & $0.472$  & $0.535$  & $0.560$
\\ \hline
$\delta_{1\pi}$    & $-0.027$ & $-0.056$ & $-0.080$
          & $-0.039$ & $-0.080$ & $-0.112$
\\ \hline
$\delta_{2\pi}$    & $0.068$  & $0.120$  & $0.160$
          & $0.091$  & $0.162$  & $0.216$
\\ \hline
$\delta_{\rm{loop}}$   & $0.041$  & $0.063$  & $0.080$
          & $0.052$  & $0.081$  & $0.104$
\\ \hline
$\delta_{\rm{mec}}$  & $0.449$  & $0.522$  & $0.560$
          & $0.524$  & $0.616$  & $0.663$
\\ \hline\hline
$\eta_{\rm{loop}}$
          & 10.2 \%  & 13.8 \%  & 16.6 \%
          & 11.0 \%  & 15.2 \%  & 18.5 \%
\\ \hline
\end{tabular}
\end{center}
\end{table}

In Table \ref{contr}, we show the contributions to
$\delta_{\rm{mec}}$
for the $\ff{1}{s}{1} \rightarrow \ff{0}{p}{1}$ transition in
$\Aeq{16}$
with various choices of parameters and
correlation functions.
Successive rows in the table give the soft-pion tree contribution,
the $1\pi$-loop correction, the $2\pi$-loop correction, the total
loop correction (sum of rows 4 and 5), and finally the total
meson-exchange contribution (sum of rows 3, 4 and 5).
The last row denoted by $\eta_{\rm{loop}}$
expresses the total loop correction
relative to the `tree' contribution.
This correction is generally small ranging from $10\%$ for a cut-off
of $d\simeq 0.7$ fm to $15\%$ for $d=0.5$ fm.

A short-range (SR) correlation function is necessary for
calculations such as these, but there is little guidance
as to its choice.  Inevitably therefore some model dependence
is injected here representing short-range phenomenology
that is under poor control. For long-range operators,
such as those arising from pion-range tree graphs, this is not a
serious
problem as the matrix elements are not strongly influenced by
the choice of correlation function.  For the operators from the
shorter ranged loop graphs, however, we can get a factor of two
difference depending on the choice.  This is evident from the
results given in Table 2.  Any conclusions, therefore,
have to be tempered by this reality.

A SR correlation function is required in this work for
two reasons, and a common function is used here to cover both.
First, there are correlations in the nuclear physics many-body
problem.  Because the two-body operators in eq. (\ref{twobd})
are expressed in terms of the relative separation of two
nucleons, $r= |\vec{r}_1-\vec{r}_2 |$,
it is convenient if the shell-model
wavefunctions are expressed in terms of similar co-ordinates.
This is quite practicable if the shell-model Hamiltonian
is the harmonic oscillator.  Then the coefficients of transformation
from a single-particle basis to relative and center-of-mass
basis are known.  However, there is one drawback to the harmonic
oscillator Hamiltonian:
its eigenfunctions are not the eigenfunctions
of the nucleon-nucleon interaction.
This is particularly important in the relative coordinate,
where the nucleon-nucleon
interaction is known to have a strong short-range repulsion
that makes the relative wavefunction go rapidly to zero as
$ r \rightarrow 0$, more rapidly than given by uncorrelated
oscillator functions.  Thus to incorporate this piece of
many-body physics it is quite common to modify two-body
operators by multiplying them by a SR correlation
function, $ {\hat g}(r)$, where $ {\hat g}(r)$ is some function
that tends to zero as $ r \rightarrow 0$ and tends to unity for
large $r$.

Second, in ChPT, loop integrations introduce high-momentum
components.
Although these loop integrals have been regularized and are
finite, there are numerical difficulties in the Fourier transform
to coordinate space.  One way to deal with this is to impose a
high-momentum cut-off.
Another way is to impose a SR correlation
function in co-ordinate space that would kill or ameliorate
divergences at the origin.  Consider a loop integral of order $L$.
{}From the counting rules of ChPT and the nature of the loop
integral, the most divergent piece has a form
$$ r^k\,\left(\frac{\ln r}{r^2}\right)^{L}$$
where $k$ is a non-negative integer depending on the process.
In our case, $k=1$ and $L=1$.
But, it is natural to
assume that the correlation function is universal for all $L$.
Thus, in order to ameliorate the divergence in all orders,
we should have
$$\lim_{r\rightarrow 0}\, r^k\,\left(\frac{\ln r}{r^2}\right)^{L}
{\hat g}(r) = 0 \ \mbox{or\ finite}$$
for any finite $L$. This is a rather severe restriction.
In practice,
because we are interested in one-loop accuracy,
this restriction can be milder.
In our case, we should have, at least,
$$\lim_{r\rightarrow 0} \frac{\ln r}{r}\,{\hat g}(r) = 0\
\mbox{or\ finite}.$$

The step function $\theta(r\smallminus d)$ is the simplest SR
correlation function
and its underlying assumption for ChPT is that a nucleon has a hard
core of radius $d$.
It certainly satisfies all the required properties stated above,
it is easy to implement and it has a very clear physical
interpretation.
Similarly a step function has been used in the nuclear many-body
problem as well \cite{PR155}, where the interpretation of $d$ now
relates to the range of the short-range repulsion in the
nucleon-nucleon interaction.

There are other choices of SR correlation functions in use in
nuclear physics.  For example, Brown \etal \ \cite{NPA286} choose
a form $1\smallminus j_0(q_c\, r)$,
with $q_c = 3.93 \mbox{fm}^{-1}$, where
$q_c$ has been adjusted to reproduce the dominant Fourier
components of a realistic two-body correlation function
calculated with the Reid soft-core potential.  Note that the value
of $q_c$ matches the inverse of the Compton wavelength for
the $\omega$-meson.  Thus this form of correlation function
is particularly appropriate for any potential, such as the
Bonn potential \cite{OBEPR}, where the short-range repulsion is
generated by `enhanced' meson-nucleon coupling for the
$\omega$-meson.  The Bonn potential is `softer' than the historic
hard-core potentials and this correlation function is
likewise softer than the form $\theta(r\smallminus d)$.  If we define
an effective cut-off,
$d_{\mbox{eff}}$,
for ${\hat g}_3(r)$ = $1\smallminus j_0(q_c\, r)$,
such that the correlation function
$\theta(r\smallminus d_{\mbox{eff}})$ gives approximately
the same result as ${\hat g}_3(r)$, then we obtain
\be
d_{\mbox{eff}}\, \simeq 0.4\ \mbox{fm}.
\ee
This value is smaller than the choices of $d_1 \simeq 0.7$ fm and
$d_2 = 0.5$ fm used with ${\hat g}_1(r)$ and ${\hat g}_2(r)$,
respectively.

The problem with the correlation function
$1\smallminus j_0(q_c\, r)$ is that
it does not meet the requirements from ChPT of ameliorating
the divergences at the origin.  In particular it could not be used for
two-loop or higher calculations.  Thus in the following discussions
we will give less weight to calculations using ${\hat g}_3(r)$.

\begin{table}[t]
\begin{center}
\caption{Contributions to $\delta_{\rm{mec}}$ with PMR
parameters and
with ${\hat g}(r)= {\hat g}_1(r)$
\label{cpmr1}}
\vskip 1mm
\begin{tabular}{|c||r|r|r|r|r|r|r|}\hline
& $\Aeq{16}$ & $\Aeq{40}$ & $\Aeq{48}$ & $\Aeq{96}$ &
$\Aeq{132}$ &
 $\Aeq{208}$ & $\Aeq{208}$\\
$\hbar\omega(\mbox{MeV})$ & 13.921 & 11.021 & 10.489 & 8.635
& 7.874
&6.883 & 6.883
\\ \hline
initial
& $\ff{1}{s}{1}$
& $\ff{1}{p}{3}$
& $\ff{1}{p}{3}$
& $\ff{2}{s}{1}$
& $\ff{1}{f}{7}$
& $\ff{1}{g}{9}$
& $\ff{2}{p}{1}$\\
final & $\ff{0}{p}{1}$ & $\ff{0}{d}{3}$ & $\ff{0}{d}{3}$
& $\ff{1}{p}{1}$ & $\ff{0}{g}{7}$ & $\ff{0}{h}{9}$
& $\ff{2}{s}{1}$ \\ \hline\hline
$\delta_{\rm{tree}}$    & $0.407$  & $0.435$  & $0.523$
          & $0.441$  & $0.554$  & $0.553$
          & $0.462$
\\ \hline
$\delta_{1\pi}$    & $-0.027$ & $-0.031$ & $-0.032$
          & $-0.028$ & $-0.034$ & $-0.035$
          & $-0.026$
\\ \hline
$\delta_{2\pi}$    & $0.068$  & $0.077$  & $0.073$
          & $0.069$  & $0.080$  & $0.084$
          & $0.066$
\\ \hline
$\delta_{\rm{loop}}$    & $0.041$  & $0.046$  & $0.040$
          & $0.041$  & $0.046$  & $0.049$
          & $0.040$
\\ \hline
$\delta_{\rm{mec}}$ & $0.449$  & $0.481$  & $0.564$
          & $0.481$  & $0.600$  & $0.602$
          & $0.502$
\\ \hline\hline
$\eta_{\rm{loop}}$
          & 10.2 \%  & 10.5 \%  &  7.7 \%
          &  9.3 \%  &  8.3 \%  &  8.8 \%
          &  8.6 \%
\\ \hline
\end{tabular}
\end{center}
\end{table}

In Table \ref{cpmr1}, we display the contributions to
$\delta_{\rm{mec}}$ for various odd-mass nuclei of
closed-shell-plus-one configuration,
where the mass number of
the closed shell, $A$, is indicated.
The purpose is to investigate the mass
dependence of the two-body contributions.
Here, the PMR parameter set
with ${\hat g}_1(r)$ is adopted.  In all cases, the $1\pi$-loop
and $2\pi$-loop corrections are small and of opposite sign
so that
the resultant $\delta_{\rm{loop}}$ is small.  Further, when
expressed relative to the `tree' contribution, the
$\eta_{\rm{loop}}$ shows little mass or state dependence.
We will discuss mass dependence further in the next section.

In Table \ref{calc1},
we present our results for $\delta_{\rm{mec}}$
and $\eta_{\rm{loop}}$
for various single-particle transitions in a
number of different nuclei ranging from light to heavy,
with the PMR parameters and the
${\hat g}_1(r)$ correlation function.

\clearpage

\begin{table}[t]
\begin{center}
\caption{Calculations of $\delta_{\rm{mec}}$ and
$\eta_{\rm{loop}}$
for a range of transitions in various nuclei
with the PMR parameters
and including the short-range correlation function,
${\hat g}(r)= {\hat g}_1(r)= \theta(r\smallminus d_1)$
with $d_1= 0.5\ m_\pi^{-1}\,\simeq\,0.7$ fm
\label{calc1}}
\vskip 1mm
\begin{tabular}{|crr||crr|}\hline
\multicolumn{3}{|c||}{$\Aeq{16}$, $\ \hbar\omega= 13.921$} &
\multicolumn{3}{|c|}{$\Aeq{40}$, $\ \hbar\omega= 11.021$} \\
\hline
Transition & $\delta_{\rm{mec}}$ & $\eta_{\rm{loop}}$ &
Transition & $\delta_{\rm{mec}}$ & $\eta_{\rm{loop}}$ \\
\hline
$\ff{0}{d}{3}\rightarrow \ff{0}{p}{3}$ & 0.351 & 9.2 \% &
$\ff{0}{f}{5}\rightarrow \ff{0}{d}{5}$ & 0.350 & 9.4 \% \\
$\ff{1}{s}{1}\rightarrow \ff{0}{p}{1}$ & 0.449 & 10.2 \% &
$\ff{1}{p}{3}\rightarrow \ff{0}{d}{3}$ & 0.481 & 10.5 \% \\
$\ff{0}{p}{1}\rightarrow \ff{0}{s}{1}$ & 0.508 & 9.3 \% &
$\ff{1}{p}{1}\rightarrow \ff{1}{s}{1}$ & 0.389 & 10.1 \% \\
& & &
$\ff{0}{d}{3}\rightarrow \ff{0}{p}{3}$ & 0.475 & 9.4 \% \\
& & &
$\ff{1}{s}{1}\rightarrow \ff{0}{p}{1}$ & 0.536 & 10.2 \% \\
\hline\hline
\multicolumn{3}{|c||}{$\Aeq{48}$, $\ \hbar\omega= 10.489$} &
\multicolumn{3}{|c|}{$\Aeq{96}$, $\ \hbar\omega= 8.635$} \\ \hline
$\ff{0}{f}{5}\rightarrow \ff{0}{d}{5}$ & 0.409 & 8.4 \% &
$\ff{0}{g}{7}\rightarrow \ff{0}{f}{7}$ & 0.386 & 8.8 \% \\
$\ff{1}{p}{3}\rightarrow \ff{0}{d}{3}$ & 0.564 & 7.7 \% &
$\ff{1}{d}{5}\rightarrow\ff{0}{f}{5}$ & 0.568 & 8.4 \% \\
$\ff{1}{p}{1}\rightarrow\ff{1}{s}{1}$ & 0.434 & 8.8 \% &
$\ff{1}{d}{3}\rightarrow\ff{1}{p}{3}$ & 0.400 & 9.3 \% \\
$\ff{0}{d}{3}\rightarrow\ff{0}{p}{3}$ & 0.507 & 9.4 \% &
$\ff{2}{s}{1}\rightarrow\ff{1}{p}{1}$ & 0.481 & 9.3 \% \\
$\ff{1}{s}{1}\rightarrow\ff{0}{p}{1}$ & 0.600 & 8.4 \% &
$\ff{0}{f}{5}\rightarrow\ff{0}{d}{5}$ & 0.468 & 9.4 \% \\
& & &
$\ff{1}{p}{3}\rightarrow\ff{0}{d}{3}$ & 0.593 & 8.7 \% \\
& & &
$\ff{1}{p}{1}\rightarrow\ff{1}{s}{1}$  & 0.502 & 9.2 \% \\
\hline\hline
\multicolumn{3}{|c||}{$\Aeq{132}$, $\ \hbar\omega= 7.874$} &
\multicolumn{3}{|c|}{$\Aeq{208}$, $\ \hbar\omega= 6.883$} \\
\hline

$\ff{0}{h}{9} \rightarrow\ff{0}{g}{9}$ & 0.400 & 7.8 \% &
$\ff{0}{i}{11} \rightarrow\ff{0}{h}{11}$ & 0.390 & 8.2 \% \\
$\ff{1}{f}{7} \rightarrow\ff{0}{g}{7}$ & 0.600 & 8.3 \% &
$\ff{1}{g}{9} \rightarrow\ff{0}{h}{9}$ & 0.602 & 8.8 \% \\
$\ff{1}{f}{5} \rightarrow\ff{1}{d}{5}$ & 0.406 & 7.7 \% &
$\ff{1}{g}{7} \rightarrow\ff{1}{f}{7}$ & 0.395 & 8.1 \% \\
$\ff{2}{p}{3} \rightarrow\ff{1}{d}{3}$ & 0.517 & 8.3 \% &
$\ff{2}{d}{5} \rightarrow\ff{1}{f}{5}$ & 0.530 & 8.8 \% \\
$\ff{2}{p}{1} \rightarrow\ff{2}{s}{1}$ & 0.437 & 8.1 \% &
$\ff{2}{d}{3} \rightarrow\ff{2}{p}{3}$ & 0.418 & 8.5 \% \\
$\ff{0}{g}{7} \rightarrow\ff{0}{f}{7}$ & 0.482 & 8.3 \% &
$\ff{3}{s}{1} \rightarrow\ff{2}{p}{1}$ & 0.476 & 8.8 \% \\
$\ff{1}{d}{5} \rightarrow\ff{0}{f}{5}$ & 0.610 & 8.9 \% &
$\ff{0}{h}{9} \rightarrow\ff{0}{g}{9}$ & 0.461 & 8.5 \% \\
$\ff{1}{d}{3} \rightarrow\ff{1}{p}{3}$ & 0.501 & 8.1 \% &
$\ff{1}{f}{7} \rightarrow\ff{0}{g}{7}$ & 0.608 & 9.3 \% \\
$\ff{2}{s}{1} \rightarrow\ff{1}{p}{1}$ & 0.559 & 8.4 \% &
$\ff{1}{f}{5} \rightarrow\ff{1}{d}{5}$ & 0.475 & 8.4 \% \\
& & &
$\ff{2}{p}{3} \rightarrow\ff{1}{d}{3}$ & 0.561 & 8.9 \% \\
& & &
$\ff{2}{p}{1} \rightarrow\ff{2}{s}{1}$ & 0.502 & 8.6 \% \\ \hline
\end{tabular}
\end{center}
\end{table}

\clearpage

{}From these tables, two conclusions emerge:
\begin{itemize}

\item
The loop correction ($\eta_{\rm{loop}}$)
is around 10 \% with some dependence
on the choice of SR correlation function.
This indicates quite a small correction,
confirming the dominance of the
soft-pion-tree graph
and the conclusions made in \cite{PMR},
where a simple-minded Fermi-gas model was adopted.
The chiral filtering conjecture appears to hold.

\item
The loop correction ($\eta_{\rm{loop}}$) is essentially
nuclear-mass and state independent.
This is easily understood in that the
tensor structure is common to the `tree', $1\pi$ and part of the
$2\pi$ operators displayed in eq.(\ref{twobd}).  These operators
only differ in their radial functions.  Furthermore
in light nuclei with LS closed shells, the other part of
the $2\pi$ operator also has the same tensor structure
because the matrix element of
$(\vec{\sigma}_1 + \vec{\sigma}_2)\,
 (\vec{\tau}_1 \times \vec{\tau}_2)$ is
the same as that of
$(\vec{\sigma}_1 \times \vec{\sigma}_2)\,
 (\vec{\tau}_1 + \vec{\tau}_2)$ in this case.
{}From Tables~3 and 4, we observe that $\eta_{\rm{loop}}$
of neutron-rich nuclei is smaller than that of $N=Z$ nuclei
by 10 or 20 \%, which may be viewed as the difference
in the matrix element of the two operators
for neutron-excess orbitals.

\end{itemize}

\section{Discussion}
\indent\indent
We should like to conclude with a more detailed discussion of the
mass dependence in the results.
To this end it is convenient to
define a ratio, $r$, where
\be
r = \frac{\delta_{\rm{mec}}(\Pb)}
{\delta_{\rm{mec}}(\Ox)} . \label{eq:r}
\ee
Here $\delta_{\rm{mec}}$ is the correction expressed as a fraction
of the $1$-body impulse-approximation matrix element.
For light nuclei in the vicinity of the closed shell,
$\Aeq{16}$, we consider the transition,
$\ff{1}{s}{1} \rightarrow \ff{0}{p}{1}$, that dominates the
spectroscopy in this mass region, while for heavy nuclei in the
vicinity of the closed shell, $\Aeq{208}$, we consider the
$\ff{1}{g}{9}
\rightarrow \ff{0}{h}{9}$ transition.
This ratio was introduced by Towner \cite{A542}, because
it is rather insensitive to the choice of SR correlation functions
and the parameters used.
Since this choice represents one of the biggest uncertainties
in the present work, the ratio, $r$, is an attractive quantity
to discuss.

There is some experimental information on the ratio, $r$.
It derives principally from experimental data
on first-forbidden beta decays,
as analyzed in the shell model by Warburton \etal\ \cite{WIAN}.
The method is to compute beta-decay matrix elements
in impulse approximation with the best available shell-model
wavefunctions, and then allow
the matrix element of the time-like part of the axial current
to be multiplied by an enhancement factor,
$\epsilon_{\rm{mec}}$.
Note that
\be
\epsilon_{\rm{mec}} =
\frac{\langle \M_\rmbody{1} \rangle +
\langle \M_\rmbody{2} \rangle}
{\langle \M_\rmbody{1} \rangle}
=1 + \delta_{\rm{mec}} .\label{eq:eps}
\ee
A value for $\epsilon_{\rm{mec}}$
is obtained from a fit between
experiment and calculation over a number of transitions in
the mass region under study.
The principal difficulty in this
analysis is the inevitable truncation required in the model
space used in the particular shell-model calculation. Thus
Warburton computes a correction for model-space truncations
to first order in perturbation theory.
This correction, however,
is dependent on the choice of residual interactions used in
the calculation.  In particular, it depends quite sensitively
on the strength of its tensor component.
For weak tensor forces, such as obtained in the Bonn
interaction \cite{OBEPR}, the value of $r$ is $1.30 \pm 0.09$,
while for strong tensor forces, such as obtained with the
Paris potential \cite{Paris}, the value is $1.49 \pm 0.10$.
In either case there is more enhancement of the axial-charge
matrix element required in the lead region than in the oxygen
region.

\begin{table}[t]
\begin{center}
\caption {Table of $r$-values.
The asterisks on the entries
are a reminder of the special meaning
of $\delta_{\rm{mec}}$, explained in the text
\label{rvalu}}
\vskip 1mm
\begin{tabular}{|clllc|}
\hline
\multicolumn{5}{|c|}{  } \\
 & $\!\!\delta_{\rm{mec}}(\Ox)$ &
 $\!\delta_{\rm{mec}}(\Pb)$ & $\ r$ &
comments \\
\multicolumn{5}{|c|}{  } \\
\hline
tree &\ \ \ 0.407 &\ \ \ 0.553 & 1.36 & table 3 line 1, present work \\
loop &\ \ \ 0.041 &\ \ \ 0.049 & 1.20 & table 3 line 4, present work \\
mec  &\ \ \ 0.449 &\ \ \ 0.602 & 1.34 & table 3 line 5, present work \\
\hline
tree &\ \ \ 0.560 &\ \ \ 0.759 & 1.36 & table 2, Towner \cite{A542} \\
correction &\ \ \ 0.031 &\ \ \ 0.055 & 1.77 & table 2, Towner \cite{A542} \\
mec  &\ \ \ 0.591 &\ \ \ 0.814 & 1.38 & table 2, Towner \cite{A542} \\
\hline
BR scaling I&\ \ \ 0.567$^*$ &\ \ \ 0.885$^*$  & 1.56 &
  Kubodera-Rho \cite{kr,ELAF93} \\
BR scaling II&\ \ \ 0.595$^*$  &\ \ \ 0.885$^*$  & 1.49& \\
\hline
Experiment &\ \ \ 0.61(3) &\ \ \ 0.79(4) & 1.30(9) & weak tensor
  force, Warburton \cite{WIAN} \\
Experiment &\ \ \ 0.61(3) &\ \ \ 0.91(5) & 1.49(10) & strong tensor
  force, Warburton \cite{WIAN} \\
\hline
\end{tabular}
\end{center}
\end{table}

In Table \ref{rvalu} we list some calculated $r$-values from the
present work and other sources \cite{A542,kr}.  The first row
gives just the one-pion tree-graph contribution, where a value
of $r = 1.36$ indicates that most of the experimental mass
dependence is accommodated by the shell model with just
the soft-pion $2$-body operator.  The mass dependence arises
because the number of core orbitals being summed over
in eq.(\ref{reduce}) is changing from light to heavy nuclei and the
oscillator frequency parameter is reducing to reflect the increase
in nuclear size.  One can easily get similar effects in Fermi
gas models.  For example, Delorme \cite{del82} was the first to
write down the expression for $\delta_{\rm{tree}}$
for a nucleon in a Fermi gas
\be
\delta_{\rm{tree}} =
\frac{\mN \kF}{4\pi^2 \F^2}
\left[ 1 + 2x - 2x(1+x)\,\ln\!\left(1+\frac{1}{x}\right) \right]
\label{fgas}
\ee
where $x = m_\pi^2/(4\kF^2)$ and $\kF$ is the Fermi momentum.
In Fermi gas models, $\kF$ is related to the nuclear density,
$\rho = 2\kF^3/(3\pi^2)$.
If it is assumed that a
valence nucleon in a light nucleus such as oxygen experiences
only one half the nuclear matter density, then the appropriate
value of $\kF$ to use in eq.(\ref{fgas}) is
$\kF = 1.08~\mbox{fm}^{-1}$.
For a valence nucleon in a heavy nucleus such as lead
the appropriate nuclear matter value of
$\kF = 1.36~\mbox{fm}^{-1}$ would be used.
Then, with the PMR parameter set,
$\delta_{\rm{mec}}(\Pb) = 0.54$
and $\delta_{\rm{mec}}(\Ox) = 0.39$
and an $r$-value of $r = 1.4$ is trivially obtained.\footnote{
With the OBEPR parameter set, these values become
$\delta_{\rm{mec}}(\Pb) = 0.63$,
$\delta_{\rm{mec}}(\Ox) = 0.45$
and $r = 1.4$. The $r$-value is insensitive to the
parameter set used, which can be understood by noting
that the $r$-value depends only on the value of $m_\pi$.}

A similar idea has been discussed by Kubodera and Rho \cite{kr}.
They argue from \cite{BR} that incorporation of approximate chiral
and scale invariances of QCD leads to a chiral Lagrangian of
low-energy hadrons in which the pion-decay constant
and the hadron masses scale universally
as a function of the matter density $\rho$ according to:
\be
\frac{\mN^{\ast}}{\mN}
\approx \frac{m^{\ast}_{\sigma}}{m_{\sigma}}
\approx \frac{m^{\ast}_{\rho}}{m_{\rho}}
\approx \frac{m^{\ast}_{\omega}}{m_{\omega}}
\approx \frac{\F^{\ast}}{\F}
\equiv \Phi(\rho)
\label{BRscale}
\ee
while the axial-coupling constant does not scale,
\be
\frac{\gA^\ast}{\gA} \approx 1.
\label{gAscale}\ee
Here the asterisk refers to a value in a nuclear medium as opposed
to the free hadron value.
Note that the pion-decay constant, $\F$,
and the heavy meson masses all scale
according to the same function.
However the mass of the pion,
being a Goldstone boson, is assumed
not to scale, $m_\pi^\ast\approx m_\pi$.
These assumptions are referred to as Brown-Rho (BR) scaling.
But there is a subtlety explained in detail
in Rho's lecture note \cite{ELAF93}:
the Gamow-Teller coupling constant $\gA$
scales in medium not due to the above BR scaling
but due to the short-range interactions between baryons;
as a result, {\em the constant
$\gA^\ast / \F^\ast$ associated with a pion exchange remains
constant},
\be
\frac{\gA^\ast}{\F^\ast} \approx \frac{\gA}{\F}.
\label{GTscale}
\ee
These considerations lead to
\be
\langle A^0(1\hyphen{\rm body})\rangle^\ast
=\Phi(\rho)^{-1}\,\langle A^0(1\hyphen{\rm body})\rangle,
\label{eq:1rho}
\ee
and
\be
\langle A^0({\rm mec})\rangle^\ast = \Phi(\rho)^{-1}\,
\langle A^0({\rm mec})\rangle ,
\label{eq:2rho}
\ee
where the asterisked quantities
should be evaluated with the scaled parameters defined in
eqs.(\ref{BRscale}, \ref{gAscale}, \ref{GTscale}),
while un-asterisked quantities should be evaluated
with the parameters in free space, given in Table~1.
Once the 1-body operator becomes $\rho$-dependent,
we need to elaborate on
the definition of $\epsilon_{\rm{mec}}$ [eq.(\ref{eq:eps})]
by specifying what 1-body matrix element
is used in the denominator.
To be consistent with the way in which
the empirical $\epsilon_{\rm{mec}}$
was deduced by Warburton \etal\, \cite{War,WIAN},
we must adopt the definition
\be
\epsilon_{\rm{mec}} \equiv
\frac{\langle A^0(\rmbody{1}) \rangle^\ast
+ \langle A^0({\rm mec}) \rangle^\ast}
{\langle A^0(\rmbody{1}) \rangle}.
\ee
The $\delta_{\rm{mec}}$ which should appear
in eq.(\ref{eq:r}) and which
for the sake of clarity is denoted here by
$\delta_{\rm{mec}}^{\rm{Warb}}$ is {\em numerically}
defined by
\be
\epsilon_{\rm{mec}}
\equiv 1 + \delta_{\rm{mec}}^{\rm{Warb}}.
\ee
Meanwhile,
eqs. (\ref{eq:1rho}) and (\ref{eq:2rho}) imply that
\be
\epsilon_{\rm{mec}}
= \Phi(\rho)^{-1}\, \left[ 1+
\frac{\langle A^0({\rm mec}) \rangle}
{\langle A^0(\rmbody{1}) \rangle}
\right] .
\ee
Reinterpreting $\delta_{\rm{mec}}$ of
eq.(\ref{eq:deltax})
in the present context, we identify
\be
\frac{\langle A^0({\rm mec}) \rangle}
{\langle A^0(\rmbody{1}) \rangle}
= \delta_{\rm{mec}}.
\ee
Then
\be
\delta_{\rm{mec}}^{\rm{Warb}} =
\Phi(\rho)^{-1}\,(1+\delta_{\rm{mec}}) -1.
\ee
We assume $\rho(\Pb)=\rho_0$ with
$\rho_0 \simeq 0.17~\mbox{fm}^{-3}$
being the normal nuclear matter density.
There is latitude in choosing $\rho(\Ox)$,
and we consider here two cases.
In Case~I we assume, as previously,
that $\rho(\Ox)= \rho_0/2$.
To illustrate the sensitivity of the results to $\rho$,
we also consider Case~II, in which
$\rho(\Ox)= 0.6\rho_0$.
A typical choice of $\Phi$ is
$\Phi(\rho) = 1 - 0.15\, (\rho/\rho_0)$.
Then, using $\delta_{\rm{mec}}$
given in Table \ref{cpmr1},
we obtain, for Case~I,
$\delta_{\rm{mec}}^{\rm{Warb}}(\Ox)=0.567$ and
$\delta_{\rm{mec}}^{\rm{Warb}}(\Pb)=0.885$,
which leads to $r=1.56$.
The corresponding numbers for Case~II are:
$\delta_{\rm{mec}}^{\rm{Warb}}(\Ox)=0.595$,
$\delta_{\rm{mec}}^{\rm{Warb}}(\Pb)=0.885$,
and $r=1.49$.
These results are listed in Table \ref{rvalu}
under the heading `BR scaling'.
We note that in this approach the scaling factor
$\Phi(\rho)$ is the principal mechanism
for explaining the observed mass dependence.

If we repeat the same procedure
using $\delta_{\rm{tree}}$ instead of
$\delta_{\rm{mec}}$, the resulting $r$-value
will be somewhat larger;
$r_{\rm{tree}}=1.59$ for Case I,
and $r_{\rm{tree}}=1.51$ for Case II.
As can be seen in Table \ref{rvalu},
the corrections coming from pion loops evaluated in ChPT
are almost nuclear mass independent,
and this explains why
the sum of tree graphs plus corrections
leads to a smaller $r$-value than the tree graphs alone.
This feature is in contrast to the results obtained by
Towner \cite{A542} in a quite different approach.
There the tree graphs are not evaluated in the soft-pion limit
but are evaluated with full momentum dependence retained,
including vertex form factors\footnote{
We remark that form factors built into the Lagrangian
have no place in ChPT;
the momentum dependence of vertices in ChPT
arises from radiative corrections,
as one calculates higher-order diagrams in chiral expansion.}.
This calculation, called the `hard-pion approach',
reduces the value of the tree graph significantly.
The reduction, however, is largely compensated
by heavy-meson pair graphs,
which also give a sizeable contribution as pointed out by
Kirchbach, Riska and Tsushima \cite{krt92}.
Thus the correction in this approach
is computed from `hard pions' plus `heavy mesons'
minus `soft pions', and is given in line 5 of Table \ref{rvalu}.
This correction has significant mass dependence, such that the
$r$-value of the sum of soft-pion tree graph plus corrections
increases relative to soft pions alone.

These two viewpoints, however, can be reconciled.  When
heavy-meson pair graphs are explicitly computed,
the largest contributions come
from $\sigma$ and $\omega$ mesons.
Mathematically
the $\sigma$-meson contribution takes the form of the $1$-body
impulse approximation with the nucleon mass replaced by an
effective mass.
This was shown by Delorme and Towner \cite{DT87}
in a different context.  Recently Birse \cite{birse} finds
a similar result in the non-topological soliton model for a nucleon
embedded in mean scalar and vector fields.
This implies that the phenomenological $\sigma$-meson
and the BR scaling play the same role,
although their chiral properties are quite different.
At this place, it should be noted that there is no way to
introduce a $\sigma$-field in a theory where chiral symmetry is
non-linearly realized. The only possible way is to define
a chiral-scalar (and of course Lorentz scalar) field
as Brown and Rho did in their BR scaling \cite{BR}.
Therefore to make comparisons with ref.\cite{A542}, BR scaling has to
be added to the ChPT results.

In conclusion, we find that corrections to the soft-pion tree graph
computed from $1$-$\pi$ and $2$-$\pi$ loop graphs in ChPT
are small and around 10 \%.
This lends further support to the chiral filtering
conjecture \cite{kdr78,rb81}.
Second, the mass dependence or density dependence
evident in the analysis of the experimental data by
Warburton \etal \ \cite{WIAN} has a variety of interlinking
explanations ranging from the trivial mass dependence
inherent in the shell model to the more fundamental role of
heavy mesons or BR scaling.

\subsubsection*{Acknowledgments}
\indent\indent
This work was initiated during the ECT$^*$ Workshop on
Chiral Symmetry in Hadrons and Nuclei organized
by M. Rho and W. Weise, and
we are indebted to the ECT$^*$ for assembling us
in Trento, Italy.
We are deeply obliged to M. Rho
for the suggestion that triggered the present study and
for many illuminating comments.
One of us (TSP) gratefully acknowledges the hospitality
extended to him by the University of South Carolina,
where the main part of this work was done.
His thanks are also due to Prof. D.-P. Min
for encouragement and many useful discussions.

\end{document}